\documentclass[prl,twocolumn,showpacs,amsmath,amssymb]{revtex4}

\usepackage{graphicx}
\usepackage{dcolumn}
\usepackage{bm}
\usepackage{epsfig}

\begin{document}

\newcommand{\ep}{\varepsilon}
\newcommand{\up}{\uparrow}
\newcommand{\dn}{\downarrow}
\newcommand{\vectg}[1]{\mbox{\boldmath ${#1}$}}
\newcommand{\vect}[1]{{\bf #1}}

\title{Spatial distribution of local currents of massless Dirac fermions in quantum transport through 
graphene nanoribbons}

\author{Liviu P. Z\^ arbo}
\author{Branislav K. Nikoli\' c}
\affiliation{Department of Physics and Astronomy, University
of Delaware, Newark, DE 19716-2570, USA}

\begin{abstract}
We employ the formalism of bond currents, expressed in terms of the nonequilibrium 
Green functions, to image the charge flow between two sites of the honeycomb lattice of  
graphene ribbons of few nanometers width. In sharp contrast to nonrelativistic electrons, current density 
profiles of quantum transport at energies close to the Dirac point in clean zigzag graphene nanoribbons (ZGNR) 
differs markedly from the profiles of charge density peaked at the edges due to zero-energy localized 
edge states. For transport through the lowest propagating mode induced by these edge states, edge vacancies 
do not affect current density peaked in the center of ZGNR. The long-range potential of a single 
impurity acts to reduce local current around it while concurrently increasing the current density along 
the zigzag edge, so that ZGNR conductance remains perfect $G=2e^2/h$.
\end{abstract}

\maketitle

{\em Introduction}.---The recent experimental discovery of a two-dimensional (2D) allotrope of carbon, termed 
{\em graphene}, has ushered unforeseen avenues to explore transport and interactions of 
low-dimensional electron system, build quantum-coherent carbon-based nanoelectronic devices, and 
probe high energy physics of ``charged neutrinos'' in table-top experiments~\cite{Geim2007}. 
Graphene represents one-atom-thick layer of carbon atoms tightly packed into a honeycomb 
crystal lattice whose  symmetries impose linear energy-momentum dispersion of the low-energy quasiparticles~\cite{Ando2005}. Moreover, its bipartite structure introduces an internal 
pseudospin degree of freedom  which connects  electrons and holes through chirality 
(projection of pseudospin on the direction of  motion) of opposite signs, so that the effective 
mass equation turns into the Weyl equation for massless Dirac fermions (such as neutrinos)~\cite{Ando2005}.

Relativistic energy spectrum, pseudospin, and zero gap with linearly vanishing density of states in the 
bulk graphene, have been probed in transport experiments unveiling the `chiral' quantum Hall effect, 
`minimal conductivity' at the charge neutrality (Dirac) point $E_F=0$, and weak-localization-type 
of quantum interference effects~\cite{Geim2007}. The intriguing concept of chirality, whose conservation would 
be responsible for the suppression~\cite{Ando2005} of backscattering from smooth (on the scale of the lattice 
constant) disorder and Klein tunneling~\cite{Katsnelson2006c} through high and wide electrostatic potential 
barriers, has also led to a number of  theoretical predictions~\cite{Cheianov2006} for esoteric micrometer-size graphene-based devices.

On the other hand, recent experiments on graphene wires of nanoscale width have demonstrated the existence of a gap in their energy spectrum,~\cite{Chen2007} which would allow GNR to replace semiconductor single-wall carbon nanotubes  
while allowing for an easy integration into nanoelectronic circuits via standard lithography end etching techniques. Direct STM imaging of the states localized at the edges of realistic GNR~\cite{Kobayashi2006}, as well as possible  chirality non-conserving scattering off the GNR edges, requires to examine the effects of edge-topology-dependent transverse subband structure~\cite{Wakabayashi2007,Rycerz2007}, edge states, impurities, and potential barriers in tailoring quantum transport properties of GNR-based devices.

\begin{figure}
\centerline{\psfig{file=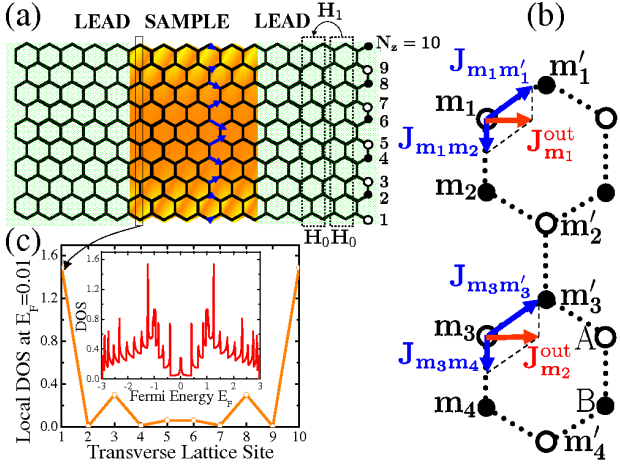,scale=0.32,angle=-90}}
\caption{(Color online) (a) Two terminal device, biased by the electrochemical potential difference $\mu_L-\mu_R=eV$, consisting of a finite ZGNR sample connected to two ideal semi-infinite ZGNR leads. The width 
of 10-ZGNR is measured by the number of zigzag chains $N_z=10$, and $a$ is the lattice constant. The definition of bond current between two sites ${\bf J}_{\bf mm'}$, shown in panel (a) by arrows connecting two supercells, and outflowing current ${\bf J}_{\bf m}^{\rm out}$ at site ${\bf m}$ is illustrated in panel (b). Panel (a) also plots the pattern of the local density of states at $E_F=0.01t$, which is dominated by the zero-energy localized edge states, as shown in panel (c) and in the corresponding total density of states in its inset.}\label{fig:zgnr}
\end{figure}

The effect of zero-energy quantum states localized at the edges of ZGNR shown in Fig.~\ref{fig:zgnr} (which originate from the gauge field generated by lattice deformation~\cite{Sasaki2006b} and reflect the topological order in the bulk of bipartite honeycomb lattice~\cite{Ryu2002}), as well as the energy gaps in armchair graphene nanoribbons (AGNR) controllable by their width, have been studied theoretically for more than a decade in equilibrium~\cite{Fujita1996,Son2006a} and conduction properties~\cite{Wakabayashi2007,Rycerz2007,Wakabayashi2000a,Peres2006b,Munoz-Rojas2006,Shi2006}. However, very 
little is known about local features of transport through GNR. Furthermore, the application of recently advanced scanning probe techniques, developed  to image local charge flow in quantum transport through 2D electron gases buried inside semiconductor heterostructures~\cite{Topinka2003}, to graphene samples is eagerly awaited~\cite{Geim2007}.  Many interesting findings are anticipated~\cite{Geim2007} when 2D electron states exposed on graphene surface are directly accessed by tunneling and local probes. Also, such transport experiments, going beyond traditional measurements of macroscopically averaged quantities, are becoming increasingly important for the development 
of nanolectronic devices---for example, recent imaging~\cite{Yoshida2007} of the charge flow in conventional $p-n$ junctions suggests that in structures shrunk below 50 nm individual positions of scarce dopants will affect their function, thereby requiring to know precisely how charge carriers propagate on the nanoscale.

Here we extend the {\em bond current} formalism for square lattices~\cite{Nikoli'c2006,Cresti2003,Nonoyama1998} to graphene honeycomb lattice, which allows us to predict spatial profiles of {\em nonequilibrium} charge and current densities. This imaging of charge flow provides direct insight into how massless Dirac fermions propagate between two neighboring lattice sites. The two-terminal device setup is shown in Fig.~\ref{fig:zgnr} where finite GNR sample is attached to two semi-infinite GNR leads. When the sample is clean, the whole structure represents infinite ZGNR (illustrated by Fig.~\ref{fig:zgnr}) or AGNR, while the disordered sample is created by introducing vacancies at its edges or short-range or long-range impurity potential within its interior. The whole structure is described by the tight-binding Hamiltonian
\begin{equation}\label{eq:tbh}
\hat{H} = -  \sum_{{\bf
mm'}} t_{\bf mm'} \hat{c}_{{\bf m}}^\dag \hat{c}_{{\bf m'}} + \sum_{{\bf m}} V_{\bf m} \hat{c}_{\bf
m}^\dag\hat{c}_{\bf m},
\end{equation}
with single $\pi$-orbital per site. Here $\hat{c}_{\bf m}^\dag$ ($\hat{c}_{\bf m}$) 
creates (annihilates) an electron in the $\pi$-orbital at the site ${\bf m}=(m_x,m_y)$, 
and $t_{\bf mm'}=t \approx 2.84$ eV is a hopping parameter between nearest neighbor 
orbitals (which allows to reproduce {\em ab initio}~\cite{Reich2002} computed structure of 
the conduction and valence bands in the vicinity of $K$ and $K'$ Dirac points located in two 
inequivalent corners of the hexagonal Brillouine zone where the bands touch conically). The impurity potential at site ${\bf m}$ is $V_{\bf m}$. 

{\it Bond current formalism on graphene honeycomb lattice}.---The central quantity of the steady-state local transport formalism on tight-binding lattices is the nonequilibrium lesser Green function $G^<_{{\bf m}'{\bf m}}(\tau=0) = \frac{i}{\hbar}\left< \hat{c}^\dag_{\bf m}\hat{c}_{\bf m} \right>  = \frac{1}{h} \int_{-\infty}^{\infty}dE G^<_{{\bf m}'{\bf m}}(E)$ function~\cite{Keldysh1965}, where $\left< \ldots \right>$ is the nonequilibrium statistical average with respect to the density matrix at time  $t^\prime=0$. It yields the magnitude of the bond current 
\begin{equation}\label{eq:bond}
J_{\bf mm'} =
\frac{2e}{h}\int\limits_{E_F-eV/2}^{E_F+eV/2} \!\! dE  \left[ t_{\bf m'm} \vect{G}^{<}_{\bf mm'}(E) - t_{\bf mm'} \vect{G}^{<}_{\bf m'm}(E) \right],
\end{equation}
between the lattice sites ${\bf m}$ and its nearest neighbor site ${\bf m'}$, and the nonequilibrium charge density at site ${\bf m}$
\begin{equation}\label{eq:charge}
n_{\bf m} = \frac{e}{2\pi i} \int\limits_{E_F-eV/2}^{E_F+eV/2} \!\! dE \, G^<_{{\bf m}{\bf m}} (E).
\end{equation}
These are the expectation values of the corresponding operators, ${J}_{\bf mm'} = \left<\hat{J}_{\bf mm'} \right>$ 
and $n_{\bf m} = e \left< \hat{N}_{\bf m}\right>$, which satisfy the charge continuity equation on 
the lattice, $ e d\hat{N}_{\bf m}/dt + \sum_{\bf m'} \left(\hat{J}_{\bf mm'} - \hat{J}_{\bf m'm} \right)=0$, 
where ${\bf m'}$ are the three nearest neighbor sublattice B sites of site ${\bf m}$ belonging to sublattice 
A, and {\em vice versa}. Thus, the bond current  $J_{\bf mm'}$ can be visualized as a bundle of 
flow lines bunched together along a link joining the two sites. 

The connection between the bond current vector ${\bf J}_{\bf mm'}$, outflowing current ${\bf J}^{\rm out}_{\bf m}$ from site ${\bf m}$ and total current $I = \sum_{{\bf m}^\prime_i} |{\bf J}_{\bf mm'}|=\sum_{{\bf m}^\prime_i} J_{\bf mm'}$ is illustrated by Fig.~\ref{fig:zgnr}(b). That is, when magnitudes of all vectors  ${\bf J}_{\bf mm'}$ (where length of the arrow is proportional to  local current) on the bonds connecting supercells in Fig.~\ref{fig:zgnr}(a) are summed up to get $I$, then $G=I/V$ gives the linear response conductance for small applied voltage bias $V = 0.001 t/e$. For zero-temperature quantum transport of electrons injected at the Fermi energy $E_F=0.01t$  there is only one open conducting channel, so that $G=I/V=2e^2/h=G_Q$ ($G_Q$ is the conductance quantum) for spatial distribution of local currents within ZGNR of Fig.~\ref{fig:zgnr}(a).

The matrix ${\bf G}^<(E)$ contains information about the occupied states in the central region, and can be obtained by solving the Keldysh equation ${\bf G}^<(E) = {\bf G}^r(E) {\bm \Sigma^<}(E){\bf G}^a(E)$. In the single-particle approximation, where interactions are of the mean-eld type, this equation can be solved exactly  by evaluating the retarded ${\bf G}^r(E)=[E-{\bf H}-U_{\bf m}-{\bm \Sigma}_L^r - {\bm \Sigma}_R^r]^{-1}$ and the advanced ${\bf G}^a(E)=[{\bf G}^r(E)]^\dag$ Green function matrices. In the absence of  inelastic scattering, the retarded 
self-energies ${\bm \Sigma}_L^r(E-eV/2)$ and ${\bm \Sigma}_R^r(E+eV/2)$ introduced  by the left and the right lead~\cite{Caroli1971}, respectively, determine ${\bm \Sigma}^<(E)= -2i  [{\rm Im}\, {\bm \Sigma}_L(E-eV/2) f_L(E-eV/2) + {\rm Im}\, {\bm \Sigma}_R(E+eV/2) f_R(E+eV/2)]$  [$f_{L,R}$ are the Fermi distribution functions of the electrons injected from the macroscopic reservoirs  through the leads and ${\rm Im} \, {\bm \Sigma}_{L,R} = ({\bm \Sigma}_{L,R}^r -{\bm \Sigma}_{L,R}^a)/2i$]. The gauge invariance of measurable quantities with respect to the shift of electric potential by a constant is satisfied on the proviso that ${\bm \Sigma}_L^r$,  ${\bm \Sigma}_R^r$ depend explicitly on  the applied  bias voltage $V$ while ${\bf G}^r(E)$ has to include the electric potential landscape $U_{\bf m}$ within the sample. 

The key issue in the study of transport through nanoelectronic devices attached to graphene leads is efficient algorithm to compute the retarded surface Green function at the terminating edge of the semi-infinite GNR. For this purpose we employ the Ando algorithm~\cite{Ando1991}, which constructs this Green function from the left and right transverse propagating exact Bloch eigenmodes. The original algorithm~\cite{Ando1991}  has to be generalized~\cite{Khomyakov2005} to handle non-invertible Hamiltonian ${\bf H}_1$  connecting  supercells consisting of atoms described by the Hamiltonian ${\bf H}_0$ (see Fig.~\ref{fig:zgnr} for illustration). 

\begin{figure}
\centerline{\psfig{file=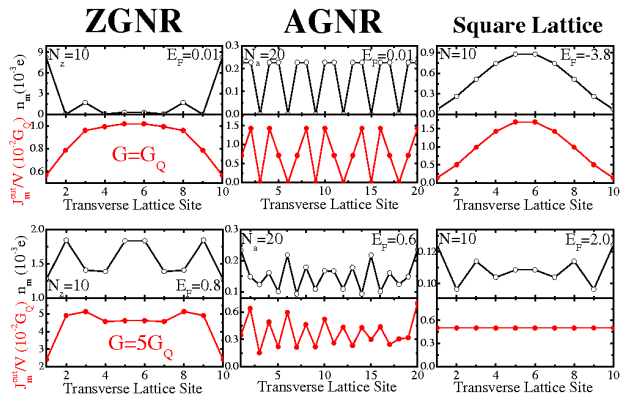,scale=0.32,angle=-90}}
\caption{(Color online) Spatial profiles of charge density $n_{\bf m}$ and local current magnitude $J^{\rm out}_{\bf m}$ at each site of the cross section of ideal (disorder and defect free) infinite ZGNRs, AGNRs, and conventional quantum 
wires defined on square tight-binding lattices. The panels in the first row correspond to zero-temperature single 
channel quantum transport with the Landauer conductance $G=G_Q$, while in the second row the Fermi energy $E_F$ is tuned to allow for multichannel transport with $G=5G_Q$ ($G_Q=2e^2/h$).}\label{fig:profiles}
\end{figure}

{\it Imaging charge flow in clean GNRs}.---In ZGNR localized edge states appear at energies close to the Fermi level $E_F=0$ of undoped graphene. Thus, they manifest as edge peaks in the local density of states (DOS)  $\rho(E,{\bf m})=-{\rm Im}\, G^r_{\bf mm}(E)/\pi$ in Fig.~\ref{fig:zgnr} at $E=0.01t$, as well as a peak in the total DOS $D(E)=\sum_{\bf m} \rho(E,{\bf m})$ at $E=0$ [where in the bulk graphene $D(E=0) \equiv 0$] shown in the same figure. They correspond to non-bonding molecular orbitals, where electrons are strongly localized near the zigzag edge composed on sublattice A sites (bottom) or sublattice B sites (top), and, therefore, cannot carry current. However, the overlap of two edge states from the top and bottom edge of a ZGNR yields bonding and anti-bonding states ensuring single conducting channel (with partially flat energy-momentum dispersion~\cite{Wakabayashi2007,Rycerz2007}) at energies arbitrarily close to the Dirac point. Therefore, ZGNR are metallic for all widths (as long as ferromagnetic ordering on zigzag edges is not taken into account~\cite{Son2006a}). This channel, together with $2n$ right moving propagating modes crossing the Fermi energy for $0<|E_F|<t$ yields odd-number conductance quantization $G(E_F)=(2n+1) G_Q$~\cite{Wakabayashi2000a,Peres2006b,Munoz-Rojas2006} in clean ZGNR. 

The Fermi energy of undoped GNR is at half-filling $E_F=0$ due to perfect electron-hole symmetry. In narrow ZGNR the gap $\Delta_{12} \sim 1/N_z$ between the second subband and $E_F=0$ is so large (e.g., $\Delta_{12}=0.4t$ for $N_z=10$) that transport would remain within the single channel regime $E_F<\Delta_{12}$ for presently achievable densities of extra carriers that can be injected into the ZGNR. Therefore, we focus on single-channel transport at $E_F=0.01t$ with maximum conductance $G=G_Q$ in the rest of the paper, while also using multichannel transport ($E_F=0.8t$) with maximum conductance $G=5G_Q$ for comparison. 

\begin{figure}
\centerline{\psfig{file=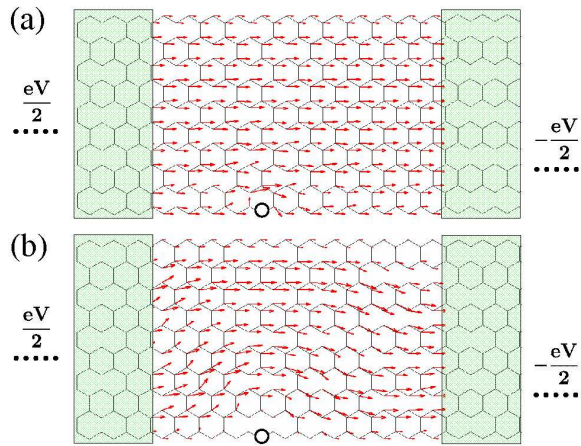,scale=0.31,angle=-90}}
\caption{(Color online)  The linear-response ($eV=0.001t$) local current ${\bf J}^{\rm out}_{\bf m}$ at each site of 10-ZGNR with single vacancy on the bottom zigzag edge. The current is proportional to the length of the arrow. 
In (a) the electrons are injected through a single conducting channel at $E_F=0.01t$, while in (b) the Fermi energy of electrons injected through ZGNR semi-infinite leads is set to $E_F=0.8$ at which there are five open channels.}\label{fig:vacancy}
\end{figure}

Figure~\ref{fig:profiles} contrasts spatial profiles of $J_{\bf m}^{\rm out}$ and $n_{\bf m}$ for single channel-transport through 10-ZGNR, 20-AGNR, and quantum wire modeled on a conventional square tight-binding lattice with 10 sites (hosting single $s$-orbitals) over its cross section. Following previous convention, the width of $N_z$-ZGNR is measured by the number of zigzag chains $N_z$, while $N_a$-AGNR have $N_a$ dimer lines across the ribbon width. In Fig.~\ref{fig:profiles}, 10-ZGNR corresponds to 20-AGNR and square lattice wires with N=10 sites per cross section in the sense that all three quantum wires support maximum of 10 open conducting channels. 

The charge density (Fig.~\ref{fig:profiles}) of low energy states is proportional to $n(y) \propto |\phi_A|^2 + |\phi_B|^2$, which in ZGNR (Fig.~\ref{fig:zgnr}) is peaked at its edges. On the other hand, the current density $j_x(y) \propto v(\Phi^\dagger \sigma_x \Phi)$ [$\sigma_x$ is the Pauli matrix representing the $x$-component of the pseudospin operator acting on the AB space] of  Dirac fermions, which are described in the low-energy (i.e., long wavelength) limit by continuum theory for the two-component wave functions $\Phi^\dagger=(\phi^*_A  \ \phi^*_B)$ [defining relative contribution of the $A$ and $B$ sublattice in the make-up of quasiparticles], reaches  maximum in the center of the ribbon. This, together with imaging of local charge flow ${\bf J}^{\rm out}_{\bf m}$ shown in Fig.~\ref{fig:vacancy}, explains why isolated edge vacancies have very little effect on the Landauer conductance $G \approx 0.98G_Q$ of single-channel transport through ZGNR~\cite{Munoz-Rojas2006}. When more channels are open for zero-temperature quantum transport, the vacancy affects not only the local current at the zigzag edge as in Fig.~\ref{fig:vacancy}(a), but also the magnitude and the direction of ${\bf J}^{\rm out}_{\bf m}$ within the bulk of ZGNR in Fig.~\ref{fig:vacancy}(b), so that its conductance drops from $G=5G_Q$ (when the vacancy is absent) to $G \approx 4G_Q$.
\begin{figure}
\centerline{\psfig{file=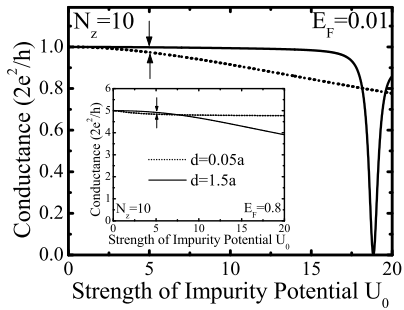,scale=0.51,angle=0}}
\caption{The Landauer conductance of ZGNR with single impurity positioned in the center of the ribbon, to generate  short-range $d=0.01a$ or long-range $d=1.5a$ Gaussian potential, as a function of the impurity potential strength. 
The electrons are injected from ZGNR lead through a single conducting channel ($E_F=0.01t$) in the main panel and 
through five conducting channels ($E_F=0.8t$) in the inset.}\label{fig:conductance}
\end{figure}

In contrast to ZGNR, spatial profiles of charge and current density in AGNR are highly inhomogeneous in both single and multichannel transport regimes. It is also instructive to compare $n_{\bf m}$ vs. $J_{\bf m}^{\rm out}$ profiles in GNRs 
with those of non-relativistic electrons in quantum wires defined on the square lattice (third column in  Fig.~\ref{fig:profiles}). Their scalar wave function $\phi(x,y)=\phi_{\rm trans}(y)e^{ikx}$ yields the charge density $n(y) \propto |\phi(y)|^2$ which has the same profile as the corresponding current density  $j_x(y) \propto \phi \partial_x \phi^* - \phi^* \partial_x \phi \propto kn(y)$. 
\begin{figure}
\centerline{\psfig{file=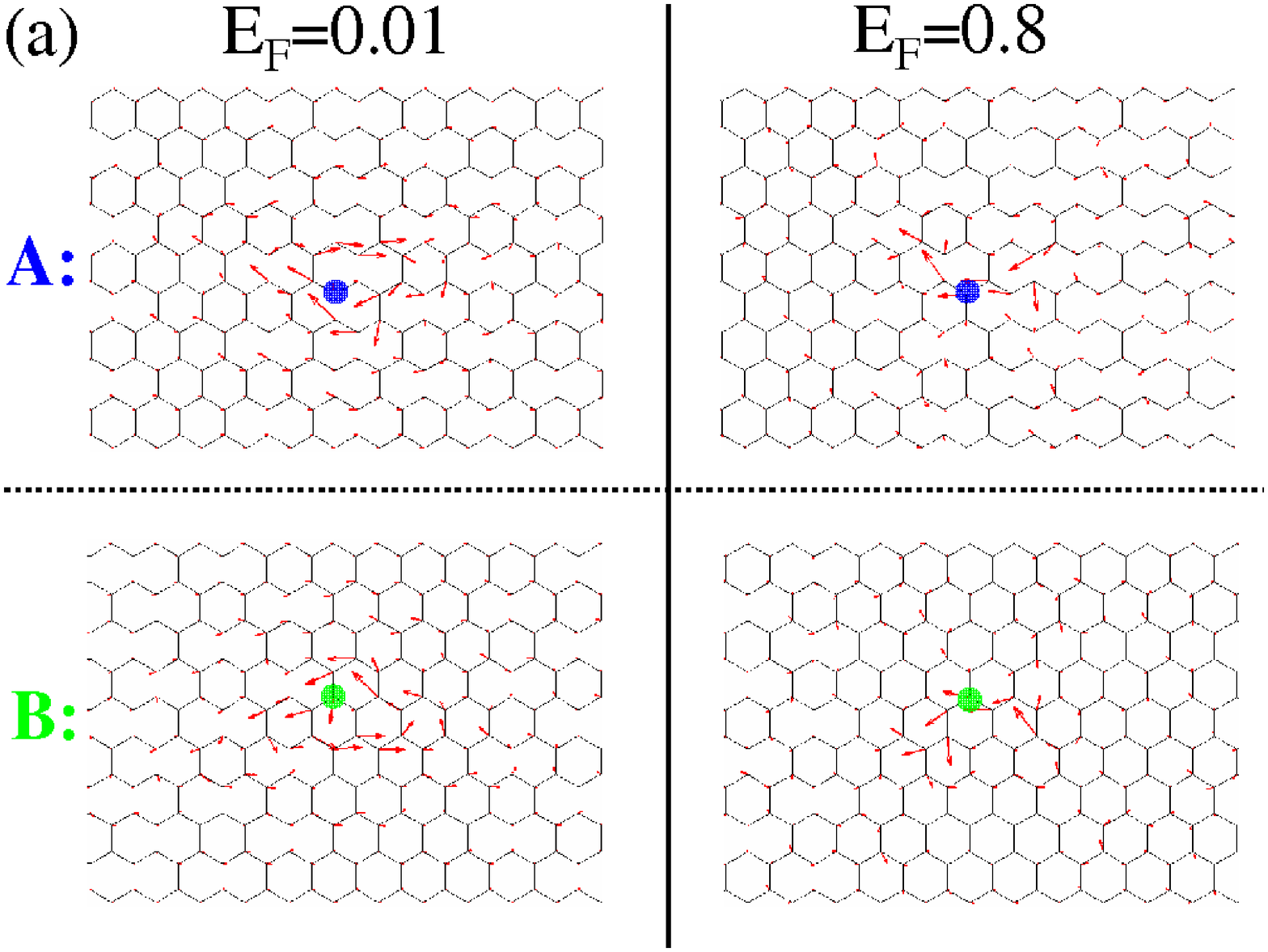,scale=0.31,angle=-90}}
\centerline{\psfig{file=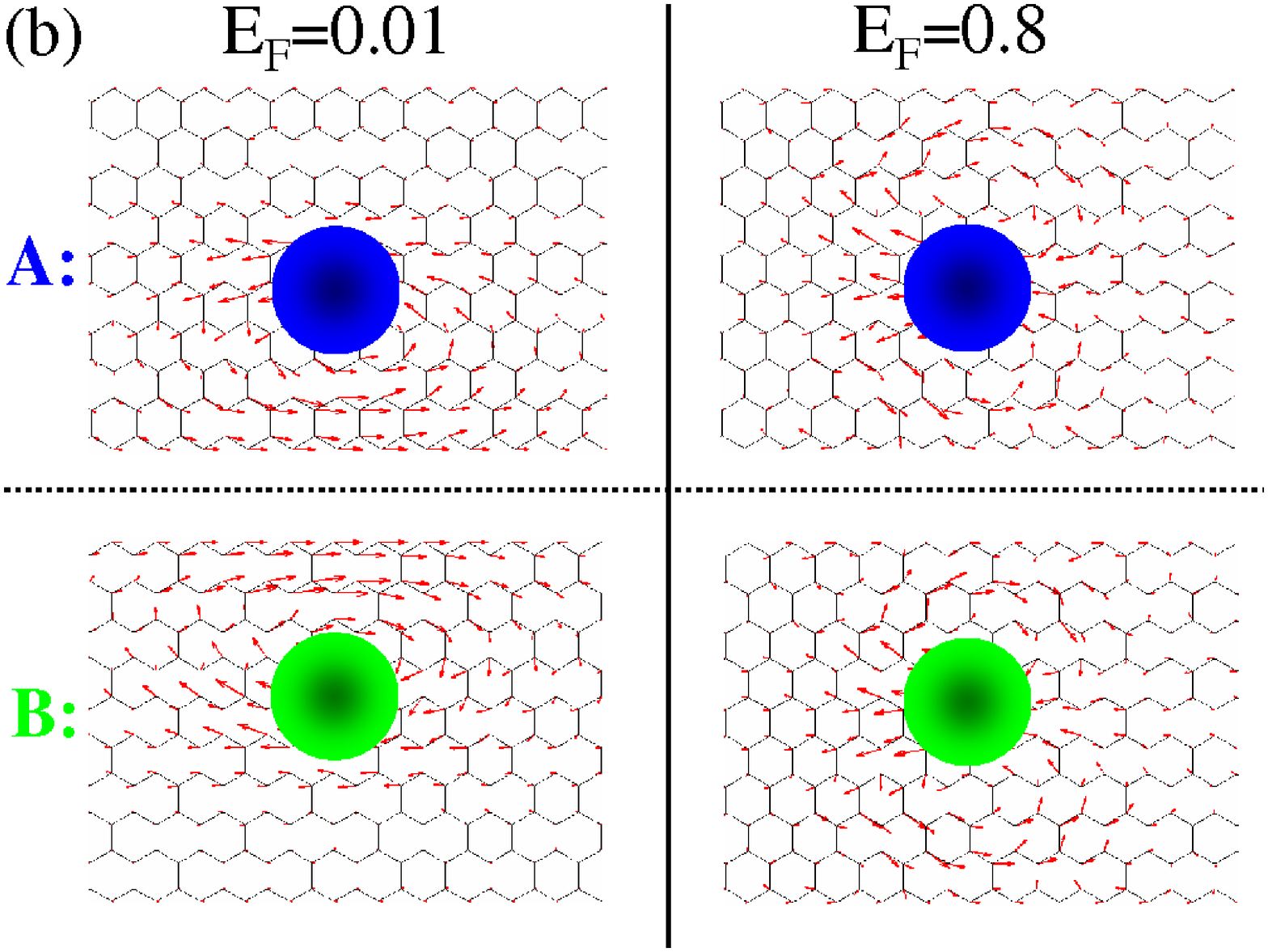,scale=0.31,angle=-90}}
\caption{(Color online) The difference ${\bf J}^{\rm out}_{\bf m}({\rm disorder})-{\bf J}^{\rm out}_{\bf m}({\rm clean})$ between the local currents at each site of a clean 10-ZGNR and the same ZGNR with disorder 
introduced as a single impurity positioned on the sublattice A or sublattice B site in its center. The 
Gaussian impurity potential of strength $U_0=5.0t$ is short-range in (a) $d=0.01a$  and long-range in (b) $d=1.5a$. The value of the corresponding conductances of single-channel ($E_F=0.01t$) and five-channel ($E_F=0.8t$) quantum transport is marked by arrows in Fig.~\ref{fig:conductance} and its inset, respectively.}\label{fig:impurity}
\end{figure}

{\it Imaging charge flow in disordered GNRs}.---In recent intense efforts to understand  experimentally observed properties of the conductivity of bulk graphene (such as its linear dependence on 
carrier concentration, minimal value $\sim e^2/h$ at the Dirac point, and absence of weak antilocalization, expected 
due to  chiral nature of electrons in graphene, or suppression of standard weak localization corrections~\cite{Geim2007}), the analysis of scattering of massless Dirac fermions from different types  of impurities has played an essential role~\cite{Ostrovsky2006}. Since energy-momentum dispersion $\varepsilon(k_x)$ of transverse propagating modes in GNRs strongly depends on the confinement effects and topology of their edges, the investigation of the disorder effects 
in transport properties of {\em mesoscopic} graphene structures requires to reexamine conditions for the absence of backscattering due to conservation of chirality~\cite{Ando2005}. For example, it has been argued recently~\cite{Wakabayashi2007} that large momentum difference between two valleys~\cite{Rycerz2007} of ZGNRs at $k_x=2\pi/3a$ and $k_x=-2\pi/3a$ (which originate from the Dirac points $K$ and $K'$ of bulk graphene) prevents inter-valley scattering by long-range disorder, so that the special conducting channel generated by the edges states remains a `chiral' propagating mode in the sense that ZGNR has perfect conductance $G(E_F)=2e^2/h$ for $|E_F| < \Delta_{12}$ that does not decay as the length of the wire is increased~\cite{Wakabayashi2007} This special channel is comprised of states belonging to only one valley (unlike higher energy modes where both valleys contribute), switching to the opposite one when changing the direction of propagation and allowing for valley filter devices~\cite{Rycerz2007}. 
The Landauer-formula-computed conductance does decay exponentially due to Anderson localization in the single or 
multichannel transport regimes when impurity potential is short-ranged~\cite{Wakabayashi2007}. 

To develop a real-space local picture of conduction through the special channel of ZGNR, we employ the Gaussian 
potential $V_{\bf m}= U \exp (-|{\bf m}-{\bf m}_0|/d^2)$ of range $d$ generated by a single impurity centered at site ${\bf m}_0$, which belongs to either sublattice A or B. The potential strength $U_0$ is defined by the normalization condition~\cite{Wakabayashi2007}, $\sum_{\bf m} U \exp (-|{\bf m}-{\bf m}_0|/d^2) = U_0$. The conductance of ZGNR as a function of $U_0$ for both short-range $d=0.05a$ (i.e., Anderson-model-type of disorder) and long-range $d=1.5a$ potential (which is short-ranged on the scale of the system size but varies smoothly on the atomic scale, corresponding to, e.g., screened charges in the substrate) is shown in Fig.~\ref{fig:conductance} in the single-channel ($E_F=0.01t$) and multichannel ($E_F=0.8t$) quantum-coherent transport regime. For long-range  impurities, the conductance remains perfect $G=2e^2/h$ up to $U_0 \sim 0.2t$ (and $G \approx 1.99 e^2/h$ up to $U_0 \sim 8t$), which is comparable to the energy difference between the transverse subbands.

The corresponding images of local charge flow through ZGNR are shown in Fig.~\ref{fig:impurity} by plotting the difference between local currents at each site ${\bf J}^{\rm out}_{\bf m}({\rm disorder})-{\bf J}^{\rm out}_{\bf m}({\rm clean})$, where $U_0=0$ in the clean case and $U_0=5t$ in the disordered ZGNR. The arrows of local currents pointing 
to the right in Fig.~\ref{fig:impurity} signify the reduction of current density due to the presence of impurity. While their sum in Fig.~\ref{fig:impurity}(a) indeed explain how current density gets reduced around the short-range impurity, in the case of long-range impurity the same reduction of current density around the impurity is compensated by the increase of current density along the zigzag edge displayed in Fig.~\ref{fig:impurity}(b). Moreover, the edge being exploited by  massless Dirac fermions to resist current degradation consists of the same sublattice sites as is the site on which the impurity is located. 

{\it Conclusions}.---We have shown how to adapt the bond current formalism to graphene honeycomb lattice which makes it 
possible to predict spatial profiles of local currents of massless Dirac fermions between two sites of the lattice. The profiles we obtain for graphene nanoribbons suggest several interesting experiments that are within the reach of present local probe-based direct charge imaging techniques~\cite{Topinka2003}: (i) in ZGNR with localized edge states the large part of current flows through the center of the ribbon which makes single-channel quantum transport largely insensitive to edge vacancies; (ii) while both short-range and long-range impurities reduce current density in the region of their influence, in the single-channel transport through the lowest transverse propagating mode generated by the edge states of ZGNR this reduction in the case of long-range impurities can be compensated by the increase of current density along the zigzag edge ensuring perfectly conducting channel $G=2e^2/h$ even in the presence of disorder.

\begin{acknowledgments}
We thank  E. Andrei, A. H. MacDonald, and P. Kim for valuable discussions.
\end{acknowledgments}

\end{document}